\documentstyle[12pt,thmsa,a4,sw20lart]{article}

\input tcilatex
\begin{document}

\begin{center}
{\bf Properties of the static, spherically symmetric solutions in the
Jordan, Brans-Dicke theory.}

{\bf \ }

S.M.KOZYREV

e-mail: Kozyrev@e-centr.ru

{\bf ABSTRACT. }
\end{center}

We have studied the properties of the static, spherically symmetric solution
of Jordan, Brans-Dicke theory. An exact interior solution for standard
space-time line element in the Schwarzschild form is obtained.

\begin{center}
{\bf 1. INTRODUCTION. }
\end{center}

Scalar-tensor and vector-tensor gravity theories provide the most natural
generalizations of General Relativity by introducing additional fields. In
these theories of gravity, the field equations are even more complex then in
General Relativity. We restrict our discussion to the Jordan-Brans-Dicke
(JBD) theory [1, 2] which among of all the alternative theories of classical
Einstein's gravity, is the most studied and hence the best known. This
theory can be thought of as a minimal extension of general relativity
designed to properly accommodate both Mach's principle [3] and Dirac's large
number hypothesis [3]. Namely, the theory employs the viewpoint in which the
scalar potential is the analog of gravitational permittivity is allowed to
vary with space and time which defined using Newton's gravitational constant
as {\it G} = 1/$\phi $. JBD theory contains a massless scalar field $\phi $
and a dimensionless constant $\omega $ that describes the strength of the
coupling between $\phi $ and the matter. The field equations of the JBD
theory are obtained by the similar variational method as the Einstein
theory, and are given as following:

\begin{center}
\begin{equation}
R_{\mu \nu }-\frac 12Rg_{\mu \nu }=\frac{8\pi }\phi T_{\mu \nu }-\frac
\omega {\phi ^2}\left( \phi _{,\mu }\phi _{,\nu }-\frac 12g_{\mu \nu }\phi
_{,\lambda }\phi ^{,\lambda }\right) -\frac 1\phi \left( \phi _{,\mu ;\nu
}-g_{\mu \nu }\phi _{;\lambda }^{;\lambda }\right) ,  \label{1}
\end{equation}

\begin{equation}
\phi _{;\lambda }^{;\lambda }=-\frac{8\pi }{3+2\omega }T,  \label{2}
\end{equation}
\end{center}

where {\it T}$_{\mu \nu }$ is the energy-momentum tensor of matter and $%
\omega $ is the coupling parameter of the scalar field. 

It is usually believed that the post-Newtonian expansions of JBD theory
reduces to general relativity when the JBD parameter \TEXTsymbol{\vert}$%
\omega $\TEXTsymbol{\vert}$\rightarrow \infty $\ (see e.g. Ref. [4]), thus
the field equations of gravitation coincide completely with those of general
relativity by replacing $\phi $ with Newton's gravitational constant {\it G} 
$\equiv $ 1/$\phi $. The JBD field f is believed to exhibit the asymptotic
behavior

\begin{center}
\begin{equation}
\phi =\phi _0+\bigcirc \left( \frac 1\omega \right) ,  \label{3}
\end{equation}
\end{center}

(where $\phi _0$ is a constant). 

However, the standard assumption the JBD solutions with {\it T} = 0
generically fail to reduce to the corresponding solutions of general
relativity when $\omega $ $\rightarrow \infty $; a number of exact JBD
solutions have been reported not to tend to the corresponding general
relativity solutions [5], [6], [7]. Moreover, some authors [8], [9], [10]
reported that the asymptotic behavior of the JBD field is not (3) when the
energy-momentum tensor {\it T}={\it T}$_\mu ^\mu $ vanishes. In this case 
{\it T} = 0, asymptotic behavior of the scalar field becomes

\begin{center}
\begin{equation}
\phi =\phi _0+\bigcirc \left( \frac 1{\sqrt{\omega }}\right) .  \label{4}
\end{equation}
\end{center}

These situation are alarming since the standard belief that JBD theory
always reduces to general relativity in the large $\omega $ limit is the
basis for setting lower limits on the $\omega $-parameter using celestial
mechanics experiments [11]. To make the situation worse, the Hawking theorem
[12] states that the Schwarzschild metric is the only spherically symmetric
solution of the vacuum JBD field equations. The proof of this theorem goes
through the fact that the JBD scalar field $\phi $ must be constant outside
the black hole and the use of the weak energy condition.

The progress in the understanding of scalar-tensor theories of gravity is
closely connected with finding and investigation of exact solutions. Shortly
after JBD theory was proposed, Heckmann obtained parametric form of the
exact static vacuum solution to the JBD equations [1]. Later Brans [2] find
the static, spherically symmetric, vacuum solution of the JBD equations in
harmonic coordinates. However, not much study has been done to give physical
interpretations to the constant parameters appearing in the Heckmann and
Brans solutions [13].

As an example, one can consider the static, spherically symmetric, vacuum
Brans solution [2] given by

\begin{center}
\begin{equation}
ds^2=-e^{2\alpha }dt^2+e^{2\beta }\left[ dr^2+r^2\left( d\theta ^2+\sin
^2\theta d\varphi ^2\right) \right] ,  \label{5}
\end{equation}

\begin{equation}
e^{2\alpha }=\left( \frac{1-B/r}{1+B/r}\right) ^{2/\sigma },  \label{6}
\end{equation}

\begin{equation}
e^{2\beta }=\left( 1+\frac Br\right) ^4\left( \frac{1-B/r}{1+B/r}\right)
^{2\left( \sigma -C-1\right) /\sigma },  \label{7}
\end{equation}

\begin{equation}
\phi =\phi _0\left( \frac{1-B/r}{1+B/r}\right) ^{-C/\sigma },  \label{8}
\end{equation}
\end{center}

where

\begin{center}
\begin{equation}
\sigma =\left[ \left( C+1\right) ^2-C\left( 1-\frac{\omega C}2\right)
\right] ^{1/2},  \label{9}
\end{equation}

\begin{equation}
B=\frac M{2C^2\phi _0}\left( \frac{2\omega +4}{2\omega +3}\right)
^{1/2},C=-\frac 1{2\omega },  \label{10}
\end{equation}
\end{center}

and where {\it M} is the mass. This solution reduces to the Schwarzschild
solution of Einstein's theory for $\omega \rightarrow \infty $\ [14].
However, choices of the constant {\it C} different from the one in Eq. (10)
are possible, and for arbitrary values of the parameter {\it C} the solution
(5)-(10) does not reduce to the Schwarzschild solution when $\omega
\rightarrow \infty $.

The values of the parameters {\it M}, {\it C}, $\omega $ in the Brans
solution are not arbitrary; it was shown that the positivity of the tensor
mass puts bounds on {\it C} and $\sigma $ [13]. However, complete
understanding of the relationships between the parameters {\it M}, {\it C}
and $\omega $, and their respective ranges of admissible values is not yet
available.

The purpose of this paper is to present a method for finding exact solutions
to the JBD field equations for the static, spherically symmetric case.
Throughout the paper, we use the metric signature - + + +, and Latin (Greek)
indices take values 1...3 (0...3); the Riemann tensor is given in terms of
the Christoffel symbols by R$_{\mu \nu \rho }^\sigma $=$\Gamma _{\mu \rho
,\nu }^\sigma $-$\Gamma _{\nu \rho ,\mu }^\sigma $+$\Gamma _{\mu \rho
}^\alpha \Gamma _{\alpha \nu }^\sigma $-$\Gamma _{\nu \rho }^\alpha \Gamma
_{\alpha \mu }^\sigma $, the Ricci tensor is R$_{\mu \rho }\equiv $R$_{\mu
\nu \rho }^\nu $, and R = g$^{\alpha \beta }$R$_{\alpha \beta }$. We use
units in which the speed of light and Newton's constant assume the value
unity.

\begin{center}
{\bf 2. METHOD OF SOLUTION (COORDINATE AND VARIABLE CHOICES).}
\end{center}

We assume that spacetime is static; the metric and the scalar field can be
chosen such that

\begin{center}
\[
g_{\mu \nu ,0}=\frac{\partial g_{\mu \nu }}{\partial t}=0;\phi
_{,0}=0;g_{0i}=0. 
\]
\end{center}

As we have already mentioned we consider standard static and spherically
symmetric space-time with a line element in the form

\begin{center}
\begin{equation}
ds^2=-e^\psi dt^2+e^\lambda dr^2+r^2\left( d\theta ^2+\sin ^2\theta d\varphi
^2\right) .  \label{11}
\end{equation}
\end{center}

For the further simplification of the problem of solving the field equations
(1), (2) we will replace variable {\it r} by {\it r}($\psi $), then the line
element (11) take a form

\begin{center}
\begin{equation}
ds^2=-e^\psi dt^2+e^{\lambda (\psi )}r^{\prime }(\psi )^2d\psi ^2+r(\psi
)^2\left( d\theta ^2+\sin ^2\theta d\varphi ^2\right) ,  \label{12}
\end{equation}
\end{center}

and we find from equations (1), (2) that

\begin{center}
\begin{eqnarray}
R_{11} &=&\frac 14\left( \left( 1+\frac{4r^{\prime }}r\right) \lambda
^{\prime }\right) -1+\left( \frac 1{r^{\prime }}-\frac 4r\right) r^{\prime
\prime }=  \label{13} \\
&&\frac 1{2\phi ^2}\left( 2\omega \phi ^{\prime 2}+\phi \left( 2\phi
^{\prime \prime }-\frac{\phi ^{\prime }\left( r^{\prime \prime }+r^{\prime
}\lambda ^{\prime }\right) }{r^{\prime }}\right) \right) ,  \nonumber
\end{eqnarray}

\begin{equation}
R_{22}=R_{33}=1+\frac{e^{-\lambda }}{2r^{\prime }}\left( r\left( r^{\prime
}\left( \lambda ^{\prime }-1\right) -r^{\prime \prime }\right) -2r^{\prime
2}\right) =\frac{r\phi ^{\prime }}\phi e^{-\lambda },  \label{14}
\end{equation}

\begin{equation}
R_{00}=\frac{e^{\psi -\lambda }}{4r\ r^{\prime 2}}\left( 4r^{\prime
2}-r\left( r^{\prime }\left( \lambda ^{\prime }-1\right) +r^{\prime \prime
}\right) \right) =-\frac{e^{\psi -\lambda }\phi ^{\prime }}{2\phi \
r^{\prime }},  \label{15}
\end{equation}

\begin{equation}
\phi _{;\lambda }^{;\lambda }=\frac{4\phi ^{\prime }}r+\frac{\phi ^{\prime }%
}{r^{\prime }}-\frac{\lambda ^{\prime }\phi ^{\prime }}{r^{\prime }}-\frac{%
\phi ^{\prime }r^{\prime \prime }}{r^{\prime 2}}+\frac{2\phi ^{\prime \prime
}}{r^{\prime }}=0.  \label{16}
\end{equation}
\end{center}

where $\psi $ is a new variable and the primes denote derivatives with
respect to $\psi $. Making use of equation (16) in (15) they simplify to

\begin{center}
\[
\frac{\phi ^{\prime }}{2\phi }-\frac{\phi ^{\prime \prime }}{2\phi ^{\prime }%
}=0. 
\]
\end{center}

Thus we obtain

\begin{center}
\begin{equation}
\phi =const\ e^{a\psi },  \label{17}
\end{equation}
\end{center}

where {\it a} is a arbitrary constant. Using the asymptotic condition in
infinity we have const =1. In the case {\it a} = 0 one can find solution of
equations (13) - (16)

\begin{center}
\begin{equation}
r=\frac b{e^\psi -1},\qquad e^\lambda =e^{-\psi },\qquad \phi =1,  \label{18}
\end{equation}
\end{center}

that identical with the Schwarzschild solution of the Einstein theory.

When {\it a }$\neq $0 making use equations (13), (15) and (16) we eliminate $%
\lambda $', $\phi $'' and obtain for {\it r}

\begin{center}
\begin{equation}
r=c\ e{\it \ }^{-\frac{\left( 1+2a\right) \left( \psi +2b\right) }2}\sec
\left( \frac{\left( \psi +2b\right) k}2\right) ,  \label{19}
\end{equation}
\end{center}

where {\it b} and {\it c} arbitrary constants, and{\it \ k =}$\sqrt{%
-1-2a\left( 1+a\left( 2+\omega \right) \right) }$ . Finally from equation
(14) we have

\begin{center}
\begin{equation}
e^\lambda =\frac{k^2}{a-a^2\omega +\left( 1+a\left( 3+a\left( 4+\omega
\right) \right) \right) \cos \gamma -k\left( 1+2a\right) \sin \gamma },
\label{20}
\end{equation}
\end{center}

where $\gamma =\left( k\left( 2b+\psi \right) \right) .$

\begin{center}
{\bf 3. ASYMPTOTIC BEHAVIOR.}
\end{center}

We will study now the geometrical properties of the metric (19), (20) for
given values of the arbitrary parameters {\it a}, {\it b} and {\it c}.
Obviously, the metric (19) and (20) must be asymptotically flat. It is
enough to show that the metric components behave in an appropriate way at
large {\it r}-coordinate values, e.g. g$_{\mu \nu }$ = $\eta _{\mu \nu }$ +
O(1/{\it r}) as {\it r }$\rightarrow \infty $. In this case we have

\begin{center}
\[
\stackunder{\psi \rightarrow 0}{\lim \ r\left( \psi \right) }\ =\infty
;\qquad \stackunder{\psi \rightarrow 0}{\lim \ \lambda \left( \psi \right) \ 
}=0 
\]
\end{center}

then we obtain from (19), (20)

\begin{center}
\begin{equation}
c\ e^{-b\left( 1+2a\right) }\sec \left( \frac{2bk}2\right) =\infty ,
\label{21}
\end{equation}

\begin{equation}
\frac{k^2}{a-a^2\omega +\left( 1+a\left( 3+a\left( 4+\omega \right) \right)
\right) \cos \left( 2bk\right) -k\left( 1+2a\right) \sin \left( 2bk\right) }%
=1.  \label{22}
\end{equation}
\end{center}

One can see there is not solution of the system equations (21), (22). In
this case only the Schwarzschild metric (18) is the spherically symmetric
solution of vacuum JBD field equations and the JBD scalar field $\phi $ must
be constant outside the matter distributions. However, for the internal
solutions increases the interaction between scalar, and tensor fields
describing gravitation.

\begin{center}
{\bf 4. INTERIOR SOLUTIONS}
\end{center}

Solving of scalar-tensor theory equations in the presence of a matter is a
difficult task due to their complexity in the general case. In the simple
case of a perfect-fluid spherically symmetric model with equation of state 
{\it p} = $\varepsilon \rho $ mach progress has been achieved in finding
exact solutions [15], [16]. When the energy-momentum tensor is specialized
to that of a perfect fluid and equation of state chosen for very high
density we derive the relation between $\phi $ and g$_{00}$. In this case we
can express the R$_0^0$ component of equations (1) in the form:

\begin{center}
\begin{equation}
-\frac 12\left[ \sqrt{-g}\phi \left( \ln \ g_{00}\right) ^{,k}\right]
_{,k}=8\pi \left( T_0^0-\frac{\omega +1}{2\omega +3}T\right) \sqrt{-g}.
\label{23}
\end{equation}
\end{center}

Furthermore in the static case equation (2) simplifies to

\begin{center}
\begin{equation}
\left[ \sqrt{-g}\phi \left( \ln \ \phi \right) ^{,k}\right] _{,k}=\frac{8\pi
T\sqrt{-g}}{2\omega +3}.  \label{24}
\end{equation}
\end{center}

Combining equations (23), (24) and making use of equation of state {\it p} = 
$\varepsilon \rho $, we obtain

\begin{center}
\begin{equation}
\left[ \sqrt{-g}\phi \left( \ln \ \left( \frac \phi {g_{00}^{\qquad
q}}\right) \right) ^{,k}\right] _{,k}=0,  \label{25}
\end{equation}
\end{center}

where

\begin{center}
\[
q=\frac{3\varepsilon -1}{\left( 2\omega +3\right) +\left( \omega +1\right)
\left( 3\varepsilon -1\right) }.
\]
\end{center}

Equation (25) is also valid outside matter with {\it q} an arbitrary
constant. If $\phi $/g$_{00}$ tends uniformly to a limit at infinity and its
second derivatives exist everywhere, then the solutions to equation (25) is
[16]

\begin{center}
\begin{equation}
\phi =e^{a\psi },  \label{26}
\end{equation}
\end{center}

where {\it a} arbitrary constant. This standard tenet about the relation
between $\phi $ and g$_{00}$ can be false, there are a number of exact JBD
vacuum solution with a $\phi $ = const. For example, in conformity with
Hawking theorem [12] only the Schwarzschild metric is the spherically
symmetric solution of the vacuum JBD field equations. At the same time
inside the matter arise the coupling between the scalar field with gravity.
Then the relation between $\phi $ and g$_{00}$ in general case is more
complexity then (26). For example, in a special case when equation of state
is {\it p} = $\varepsilon \rho $, $\omega $ = -(1 + 6$\varepsilon $) / (1 - 3%
$\varepsilon $) and g$_{rr}$=const one can find relation between $\phi $ and
g$_{00}$ in the form:

\begin{center}
\begin{equation}
g_{00}=\phi ^{-2}\left( -\frac{C_1}r+C_2\right) ^2,  \label{27}
\end{equation}
\end{center}

where {\it C1} and {\it C2} is a arbitrary constant.

However, in the present work, we investigate only $\phi $= e$^{a\psi }$
solutions to equation (25). In the paper [15], Bruckman and Kazes
demonstrated that, in the case of cold ultrahigh-density static
configurations with relation (26) between $\phi $ and g$_{00}$, solutions to
the Brans-Dicke equations have the metric component g$_{rr}$ is a constant.
Assuming that space-time is static, spherically symmetric, and the metric is
chosen in the form (11), the relation between $\phi $ and g$_{00}$ is $\phi $%
= e$^{a\psi }$ and g$_{rr}$=1, we find static solution of equations (1), (2):

\begin{center}
\begin{equation}
\psi =c_2+\frac{2\left( 1+2a\right) \log \left( r^2\left( 1+4a^2\left(
1+\omega \right) \right) +4\left( 1+2a\right) c_1\right) }{1+4a^2\left(
1+\omega \right) }.  \label{28}
\end{equation}
\end{center}

Moreover using the relation (26) we can find solution of equations (13)-(16)
with $\rho $= 0 and {\it p}$\neq $ 0

\begin{center}
\[
e^\psi =\left( -\frac{4\pi \ r^2\left( 6+\omega \right) }{3+2\omega }\right)
^{\frac{2\left( 1+\omega \right) }{2+\omega }}, 
\]

\[
e^\lambda =\frac{6+\omega }{2+\omega }, 
\]

\[
p={\it \ }e^{-\frac \psi 2},
\]

\[
\phi =\left( -\frac{4\pi \ r^2\left( 6+\omega \right) }{3+2\omega }\right)
^{\frac 1{2+\omega }}. 
\]
\end{center}

There are not analogs of this solution in Einstein theory.

\begin{center}
{\bf 5. DISCUSSIONS }
\end{center}

Analyzing the static, spherically symmetric solution of Jordan, Brance -
Dicke theory we find the following. The Schwarzschild metric is the only
spherically symmetric solution of the vacuum JBD field equations. Moreover,
when the energy-momentum tensor {\it T}={\it T}$_\mu ^\mu $ vanishes one can
use $\phi $ = const scalar field outside the matter and the field equations
of the JBD theory become the similar as the equations of the Einstein
theory. Then in empty space there is not a difference between scalar-tensor
theories (as well as vector-metric theories [17]) and Einstein theory. In
this case in empty space celestial-mechanical experiments to reveal a
difference between scalar-tensor theories and Einstein theory is not
presented possible. However, scalar field inside the matter has
characteristics like gravitation permittivity of material similar
electromagnetic permittivity of material in Maxwell theories of
electromagnetism [18]. So, in the case of our suggestion is correct, effects
considered like the fifth force and the differences between JBD and an
Einstein's theories should be experimentally tested in substance.

\begin{center}
{\bf ACKNOWLEDGMENTS}
\end{center}

This article is dedicated to the memory of Dr. V.I.Bashkov. He was of great
help teaching me some of the interesting of scalar gravity and discussing
various alternative ideas and its role in the development of gravity
theories.

\begin{center}
{\bf REFERENCES}
\end{center}

1. P. Jordan, Schwerkraft und Weltall, Vieweg (Braunschweig) 1955.

2. C.H. Brans, Phys.Rev.125, 2194 (1962).

3. C.H. Brans, '' Gravity and the Tenacious Scalar Field'', gr-qc/9705069.

4. S. Weinberg, Gravitation and Cosmology (Wiley, New York, 1972).

5. T. Matsuda, Progr. Theor. Phys. 47, 738 (1972).

6. C. Romero and A. Barros, Gen. Rel. Grav. 25, 491 (1993).

7. M.A. Scheel, S.L. Shapiro and S.A. Teukolsky, Phys. Rev. D 51, 4236
(1995).

8. N.Banerjee, S.Sen, Phys. Rev. D56, 1334 (1997).

9. V. Faraoni, Phys. Rev. D59, 084021 (1999).

10. A. Miyazaki, gr-qc/0012104 (2000).

11. C. M. Will, Theory of experiment in Gravitational Physics (Camb. Univ.
Press, Cambridge, 1993).

12. S. W. Hawking, Commun.Math. Phys.25, 167 (1972).

13. A. Beesham, Mod. Phys. Lett. A 13, 805 (1998).

14. C. Romero and A. Barros, Phys. Lett. A 173, 243 (1993).

15. W.F.Bruckman, E.Kazes, Phys. Rev. D16, 261 (1977).

16. W.F.Bruckman, E.Kazes, Phys. Rev. D16, 269 (1977).

17. V.Bashkov, S. Kozyrev, Problems of high-energy physics and field theory,
22, Protvino, (1991).

18. V.Bashkov, S. Kozyrev, preprint gr-qc/0103009 (2001).

\end{document}